\documentclass[aps,prl,superscriptaddress,showpacs,letter]{revtex4}
\usepackage{amsfonts,amssymb,amsmath}
\usepackage{color}
\usepackage[dvips]{graphicx}
\usepackage{dcolumn}
\usepackage{bm}
\usepackage{slashbox}

\begin{document}
\title{Dynamical Evolution of Interacting Modified Chaplygin Gas}

\author{Song Li\footnote{E-mail:~~lisong819@163.com}}
  \affiliation{Department of Physics, Beijing Normal University, Beijing 100875, China}

\author{Yongge Ma\footnote{E-mail:~~mayg@bnu.edu.cn}}
  \affiliation{Department of Physics, Beijing Normal University, Beijing 100875, China}

\author{Yun Chen}
  \affiliation{Department of Astronomy, Beijing Normal University, Beijing 100875, China}

\begin{abstract}
The cosmological model of the modified Chaplygin gas interacting
with cold dark matter is studied. Our attention is focused on the
final state of universe in the model. It turns out that there exists
a stable scaling solution, which provides the possibility to
alleviate the coincidence problem. In addition, we investigate the
effect of the coupling constants $c_{1}$ and $c_{2}$ on the
dynamical evolution of this model from the statefinder viewpoint. It
is found that the coupling constants play a significant role during
the dynamical evolution of the interacting MCG model. Furthermore,
we can distinguish this interacting model from other dark energy
models in the $s-r$ plane.

\end{abstract}

 \maketitle

\section{I.~~Introduction}

There is increasing evidence that our universe is presently in a
state of cosmic accelerating expansion~\cite{phys1}. This result has
been confirmed with observations via Supernovae Ia~\cite{phys2},
Cosmic Microwave Background anisotropies~\cite{phys3}, Large Scale
Structure formation~\cite{phys4}, baryon oscillations~\cite{phys5}
and weak lensing~\cite{phys6}, etc. These observations strongly
suggest that the universe is spatially flat and is dominated by an
extra component with negative pressure, dubbed as dark
energy~\cite{phys7,phys8}. In the recent years, various candidates
of dark energy have been proposed. The simplest candidate is the
cosmological constant~\cite{phys9}. But this scenario is plagued by
the severe fine-tuning problem and the coincidence problem. Other
possible forms of dark energy include quintessence~\cite{phys10},
k-essence~\cite{phys11}, phantom~\cite{phys12}, Born-Infeld
scalars~\cite{phys13}, quintom~\cite{phys14}, tachyon
field~\cite{phys15}, holographic dark energy~\cite{phys16}, and so
on. Additionally, the conjecture that dark energy and dark matter
can be unified by using the so-called Chaplygin gas (CG) obeying an
exotic equation of state (EoS)
\begin{equation}
p_{g}=-\frac{A}{\rho_{g}}
\end{equation}
has been investigated in several literatures~\cite{phys17}.
Recently, the CG model was generalized to some possible
forms~\cite{phys18}. As an alternative model, the EoS of the
generalized Chaplygin gas (GCG)~\cite{phys19} reads
\begin{equation}
p_{g}=-\frac{A}{\rho_{g}^{\alpha}},
\end{equation}
where $0\leq\alpha\leq1$. It is clear that the case $\alpha=1$
corresponds to the CG. Within the framework of
Friedmann-Robertson-Walker (FRW) cosmology, this EoS leads, after
inserted into the relativistic energy conservation equation, to an
energy density evolving as
\begin{equation}
\rho_{g}=(A+\frac{B}{a^{3(1+\alpha)}})^{\frac{1}{1+\alpha}},
\end{equation}
where $B$ is an integration constant. Hence, we see that at early
time the energy density behaves as a dust-like matter, while at late
time it behaves like a cosmological constant. So the GCG model can
also be interpreted as an entangled mixture of dark matter and dark
energy. This dual role is at the heart of the surprising properties
of the GCG model. However, these so-called unified dark matter
models have been ruled out because they produce oscillations or
exponential blowup of the matter power spectrum inconsistent with
observation~\cite{phys69}. But no observation so far rule out the
possibility of the GCG as dark energy though it is disfavored as
dark matter. In fact, the dynamical properties of the GCG have been
studied in Ref.~\cite{phys20}, which indicates that the EoS of the
GCG may cross the so-called phantom divide $w=-1$ if there exists an
interaction between the GCG and dark matter. Therefore, it is
worthwhile to further understand such unified dark matter models by
studying another candidate for the generalization of the CG,
referred to as the modified Chaplygin gas
(MCG)~\cite{phys21,phys22}, which is characterized by a simple EoS
\begin{equation}
p_{g}=A\rho_{g}-\frac{B}{\rho_{g}^{\alpha}},
\end{equation}
where $A$, $B$ and $\alpha$ are constants and $0\leq\alpha\leq1$.
The attractive feature of this model is that the EoS looks like
that of two fluids, one obeying a perfect EoS $p=A\rho$ and the
other being the GCG. From Eq. (4), it is easy to see that the MCG
reduces to the GCG if $A=0$ and to the perfect fluid if $B=0$.
Accordingly, the evolution of the energy density is given by
\begin{equation}
\rho_{g}=(\frac{B}{1+A}+\frac{C}{a^{3(1+A)(1+\alpha)}})^{\frac{1}{1+\alpha}},
~~~~ (A\neq-1)
\end{equation}
where $C$ is the constant of integration. Thus the MCG behaves as a
radiation (when $A=1/3$) or a dust-like matter (when $A=0$) at early
stage, while as a cosmological constant at later stage.

As we all know, observations at the level of the solar system
severely constrain non-gravitational interactions of baryons,
namely, non-minimal coupling between dark energy and ordinary matter
fluids is strongly restricted by the experimental tests in the solar
systems~\cite{phys23}. However, since the nature of dark sectors
remains unknown, it is possible to have non-gravitational
interactions between dark energy and dark matter. In this paper, we
study the dynamical evolution of the interacting MCG model by
considering an interaction term between the MCG and cold dark
matter. In our scenario, we find that there exists a stable scaling
solution, which is characterized by a constant ratio of the energy
densities of the MCG and cold dark matter. This provides the
possibility to alleviate the coincidence problem. Moreover, we find
that the final state is determined by the parameters of the MCG,
$\alpha$, $A$, and the coupling constants $c_{1}$ and $c_{2}$. The
latter two parameters represent the transfer strength between the
MCG and dark matter. Interestingly, we find that the EoS of the MCG
$w_{g}$ tends to a constant, which is only determined by the
coupling constants $c_{1}$ and $c_{2}$. However, both the EoS of the
total cosmic fluid $w$ and the deceleration parameter $q$ tend to
$-1$, which are independent of the choice of values for the
parameters. This indicates that the cosmic doomsday is avoided and
the universe enters to a de Sitter phase and thus accelerates
forever. Further, we investigate the effect of the coupling
constants $c_{1}$ and $c_{2}$ on the dynamical evolution of this
model from the statefinder viewpoint. By our analysis, we see that
the coupling cantants $c_{1}$ and $c_{2}$ play a significant role
during the dynamics of the interacting MCG model. It is also
worthwhile to note that, we can distinguish this interacting model
from other dark energy models in the $s-r$ plane.

In Sec. II, we study the dynamical evolution of the interacting MCG.
Then we apply the statefinder diagnosis to the interacting modified
Chaplygin gas model for various different parameters in Sec. III.
The conclusions are summarized in Sec. IV.

\section{II.~~Dynamics analysis}

In our scenario, the universe contains the MCG $\rho_{g}$ (as dark
energy), cold dark matter $\rho_{m}$ and baryonic matter $\rho_{b}$.
For a spatially flat universe, the Friedmann equation is
\begin{equation}
H^{2}=\frac{\kappa^{2}}{3}\rho=\frac{\kappa^{2}}{3}(\rho_{g}+\rho_{m}+\rho_{b}),
\end{equation}
where $H$ is the Hubble parameter, $\kappa^{2}\equiv8\pi G$ and
$\rho$ is the total energy density (nature units $c=\hbar=1$ is used
throughout the paper). Then differentiating the above equation with
respect to cosmic time $t$ and using the total energy conservation
equation
\begin{equation}
\dot{\rho}+3H(\rho+p)=0,
\end{equation}
where $p$ is the total pressure of the background fluid, we can get
the Raychaudhuri equation
\begin{equation}
\dot{H}=-\frac{\kappa^{2}}{2}(\rho_{g}+p_{g}+\rho_{m}+p_{b}),
\end{equation}
in which $p_{g}$ represents the pressure of the MCG.

We postulate that the two dark sectors interact through the
interaction term $Q$ and the baryonic matter only interacts
gravitationally with the dark sectors. Then the continuity equation
is written as
\begin{equation}
\dot{\rho}_{b}+3H(\rho_{b})=0,
\end{equation}
\begin{equation}
\dot{\rho}_{m}+3H(\rho_{m})=Q,
\end{equation}
\begin{equation}
\dot{\rho}_{g}+3H(\rho_{g}+p_{g})=-Q.
\end{equation}
Clearly, $Q$ is the rate of the energy density exchange in the dark
sectors and the sign of $Q$ determines the direction of energy
transfer. A positive $Q$ corresponds to the transfer of energy from
dark energy to dark matter, while a negative $Q$ represents the
other way round. Due to the unknown nature of dark sectors, there is
as yet no basis in fundamental theory for a special coupling between
two dark sectors. So the interaction term $Q$ discussed currently
has to be chosen in a phenomenological way~\cite{phys40}. One
possible choice for the interaction term is
$Q=3H(c_{1}\rho_{m}+c_{2}\rho_{g})$~\cite{phys41}, where $c_{1}$ and
$c_{2}$ are coupling constants. This form was first proposed in
~\cite{phys42} and it is a more general form than those found
in~\cite{phys40,phys43}, which can be obtained when $c_{1}=c_{2}=c$,
$c_{1}=0$ or $c_{2}=0$.

\subsection{A.~~~~Stability Analysis}

To analyze the evolution of the dynamical system, we introduce the
following dimensionless variables:
\begin{equation}
x\equiv\frac{\kappa^{2}\rho_{g}}{3H^{2}},~~~~~~~~y\equiv\frac{\kappa^{2}p_{g}}{3H^{2}},~~~~~~~~z\equiv
\frac{\kappa^{2}\rho_{m}}{3H^{2}}.
\end{equation}
Accordingly, the density of the baryonic matter is determined by the
Friedmann constraint (6) as
\begin{equation}
\frac{\kappa^{2}\rho_{b}}{3H^{2}}=1-x-z,
\end{equation}
which implies that $0\leq x+z\leq1$ and $0\leq x,z\leq1$.
Furthermore, using these variables, the EoSs of the MCG and the
total cosmic fluid are respectively given by
\begin{eqnarray}
&&w_{g}=\frac{p_{g}}{\rho_{g}}=\frac{y}{x},\\
&&w=\frac{p}{\rho}=\frac{p_{g}}{\rho_{g}+\rho_{m}+\rho_{b}}=y.
\end{eqnarray}
The sound velocity and the deceleration parameter respectively read
\begin{eqnarray}
&&c^{2}_{s}=\frac{\partial
p_{g}}{\partial\rho_{g}}=-\alpha\frac{y}{x}+(1+\alpha)A,\\
&&q=-\frac{\ddot{a}a}{\dot{a}^{2}}=-1+\frac{3}{2}(1+y).
\end{eqnarray}
Note that the condition for acceleration is $w<-\frac{1}{3}$ and the
physically meaningful range of the sound velocity is $0\leq
c^{2}_{s}<1$. Using Eqs. (6)-(11), we can obtain the following
autonomous system:
\begin{eqnarray}
x^{'}&=&-3[(1+c_{2})x+y+c_{1}z]+3x(1+y),\\
y^{'}&=&-3[(1+c_{2})x+y+c_{1}z][-\alpha\frac{y}{x}+(1+\alpha)A]+3y(1+y),\\
z^{'}&=&-3[-c_{2}x+(1-c_{1})z]+3z(1+y),
\end{eqnarray}
where the prime denotes a derivative with respect to $N\equiv lna$.
We set the current scale factor by $a_{0}=1$. Then the current value
of $N$ reads $N_{0}=0$. Setting $x^{'}=y^{'}=z^{'}=0$, we can obtain
the critical points $(x_{*},y_{*},z_{*})$ of the autonomous system
as follows:
\begin{itemize}
  \item \textbf{point $(a)$}:~~~
($\frac{A-c_{1}+c_{2}}{2A}(1+x_{s})$,~~$\frac{A-c_{1}+c_{2}}{2}(1+x_{s})$,~~$1-\frac{A-c_{1}+c_{2}}{2A}(1+x_{s})$),\\
  \item \textbf{point $(b)$}:~~~
($\frac{A-c_{1}+c_{2}}{2A}(1-x_{s})$,~~$\frac{A-c_{1}+c_{2}}{2}(1-x_{s})$,~~$1-\frac{A-c_{1}+c_{2}}{2A}(1-x_{s})$),\\
  \item \textbf{point $(c)$}:~~~
($\frac{1-c_{1}}{1-c_{1}+c_{2}}$,~~$-1$,~~$\frac{c_{2}}{1-c_{1}+c_{2}}$).
\end{itemize}
Here the parameter $x_{s}$ is defined by
\begin{equation}
x_{s}=\sqrt{1+\frac{4Ac_{1}}{(A-c_{1}+c_{2})^{2}}}.
\end{equation}
Thus, we can constrain the parameters in the model under the
physically meaningful conditions, namely, $0<x_{*}\leq1$ and $0\leq
c^{2}_{s*}<1$. Concretely, we analyze the existence conditions for
these three points respectively:

$\bullet$ \textbf{points $(a)$ and $(b)$}: Considering the
physically meaningful range of $x_{*}=\frac{A-c_{1}+c_{2}}{2A}(1\pm
x_{s})$, we can obtain that $0<\frac{A-c_{1}+c_{2}}{2A}(1\pm
x_{s})\leq1, A\neq0$, and moreover, $x_{s}$ is real. Together with
the constraint of the sound velocity $c^{2}_{s*}=A$, namely, $0\leq
A<1$, we can conclude that the existence conditions of points $(a)$
and $(b)$ are $c_{1}<0$, but $c_{1}\neq -1$,
$-(1-\sqrt{-c_{1}})^{2}<c_{2}<0$ and
$(\sqrt{-c_{1}}+\sqrt{-c_{2}})^{2}\leq A<1$.

$\bullet$ \textbf{point $(c)$}: According to the meaningful range of
$x_{*}=\frac{1-c_{1}}{1-c_{1}+c_{2}}$, we know that
$0<\frac{1-c_{1}}{1-c_{1}+c_{2}}\leq1$, and therefore, $c{1}<1,
c_{2}\geq0$ or $c{1}>1, c_{2}\leq0$. Furthermore, the constraint of
the sound velocity $c^{2}_{s*}$ implies that the range of $A$ is
$-\frac{\alpha}{1+\alpha}\frac{1-c_{1}+c_{2}}{1-c_{1}}\leq
A<\frac{1}{1+\alpha}(1-\alpha\frac{1-c_{1}+c_{2}}{1-c_{1}})$. Thus,
the existence conditions of point $(c)$ are obtained. Results from
our above analysis are concluded in Table I.

\begin{table}[tbp]
\begin{center}
\begin{tabular}{|c||c|c|c|c|c|}
  \toprule
  Point & Existence & Eigenvalues & Stability & $w_{*}$ & Acceleration
\\\hline\hline
  $~$ & $c_{1}<0$, but $c_{1}\neq -1$ & $\lambda_{1}=3(A-c_{1}+c_{2})x_{s}$, & ~ & ~ &
  ~
  \\
  $(a)$ & $-(1-\sqrt{-c_{1}})^{2}<c_{2}<0$ & $\lambda_{2}=\frac{3}{2}(A-c_{1}+c_{2})(1+x_{s})$, & Unstable &
  $\frac{A-c_{1}+c_{2}}{2}(1+x_{s})$ &
  No
  \\
  $~$ & $(\sqrt{-c_{1}}+\sqrt{-c_{2}})^{2}\leq A<1$ & $\lambda_{3}=3(1+\alpha)[1+\frac{(A-c_{1}+c_{2})}{2}(1+x_{s})]$
  & ~ & ~ &
  ~
  \\\hline
  $~$ & $c_{1}<0$, but $c_{1}\neq -1$ & $\lambda_{1}=-3(A-c_{1}+c_{2})x_{s}$, & ~ & ~ &
  ~
  \\
  $(b)$ & $-(1-\sqrt{-c_{1}})^{2}<c_{2}<0$ & $\lambda_{2}=\frac{3}{2}(A-c_{1}+c_{2})(1-x_{s})$, & Saddle &
  $\frac{A-c_{1}+c_{2}}{2}(1-x_{s})$ &
  No
  \\
  $~$ & $(\sqrt{-c_{1}}+\sqrt{-c_{2}})^{2}\leq A<1$ & $\lambda_{3}=3(1+\alpha)[1+\frac{(A-c_{1}+c_{2})}{2}(1-x_{s})]$
  & ~ & ~ &
  ~
  \\\hline
  $~$ & ~ & ~ & Stable if & ~ &
  ~
  \\
  $~$ & ~ & ~ & $c_{1}<1,c_{2}\geq 0$ and & ~ &
  ~
  \\
  $~$ & $c_{1}<1,c_{2}\geq 0$ or & $\lambda_{1}=-3$, & $-\frac{\alpha}{1+\alpha}\frac{1-c_{1}+c_{2}}{1-c_{1}}\leq A$
  & ~ &
  ~
  \\
  $(c)$ & $c_{1}>1,c_{2}\leq 0$ & $\lambda_{2,3}=-\frac{3}{2(c_{1}-1)}\{-\alpha (1-c_{1}+c_{2})$ &
  $<\frac{1}{1+\alpha}(1-\alpha\frac{1-c_{1}+c_{2}}{1-c_{1}})$ & $-1$ &
  All
  \\
  $~$ & $-\frac{\alpha}{1+\alpha}\frac{1-c_{1}+c_{2}}{1-c_{1}}\leq A$ & $-(1-c_{1})[2-c_{1}+c_{2}+(1+\alpha)A]
  \pm \lambda_{s}\}$ & Saddle if & ~ &
  $\alpha, A, c_{1}$ and $c_{2}$
  \\
  $~$ & $<\frac{1}{1+\alpha}(1-\alpha\frac{1-c_{1}+c_{2}}{1-c_{1}})$ & ~ & $c_{1}>1,c_{2}\leq 0$ and & ~ &
  ~
  \\
  $~$ & ~ & ~ & $-\frac{\alpha}{1+\alpha}\frac{1-c_{1}+c_{2}}{1-c_{1}}\leq A$ & ~ &
  ~
  \\
  $~$ & ~ & ~ & $<\frac{1}{1+\alpha}(1-\alpha\frac{1-c_{1}+c_{2}}{1-c_{1}})$ & ~ &
  ~
  \\
  \botrule
\end{tabular}
\end{center}
\caption{The properties of the critical points for the interacting
modified Chaplygin gas model. Here, the parameter $\lambda_{s}$ is
defined in Eq. (25).}
\end{table}

To study the stability of the critical points for the autonomous
system, we substitute linear perturbations $x\rightarrow
x_{*}+\delta x$, $y\rightarrow y_{*}+\delta y$ and $z\rightarrow
z_{*}+\delta z$ about the critical points into the autonomous system
Eqs. (18)-(20). To first-order in the perturbations, one gets the
following evolution equations of the linear perturbations:
\begin{eqnarray}
\delta x^{'}&=&3(y_{*}-c_{2})\delta x+3(x_{*}-1)\delta y-3c_{1}\delta z,\\
\delta y^{'}&=&-3[\alpha \frac{y_{*}}{x_{*}}(\frac{y_{*}}{x_{*}}+c_{1}\frac{z_{*}}{x_{*}})+(1+c_{2})(1+\alpha)A]
\delta x \nonumber\\
&&+3[1-(1+\alpha)A+(1+c_{2})\alpha+2y_{*}+\alpha (2\frac{y_{*}}{x_{*}}+c_{1}\frac{z_{*}}{x_{*}})]\delta y-3c_{1}
[-\alpha \frac{y_{*}}{x_{*}}+(1+\alpha)A]\delta z,\\
\delta z^{'}&=&3c_{2}\delta x+3z_{*}\delta y+3(y_{*}+c_{1})\delta z.
\end{eqnarray}
The three eigenvalues of the coefficient matrix of Eqs. (22)-(24)
determine the stability of the critical points. We list the three
eigenvalues for each point in Table I. Moreover, the condition for
stability and acceleration (i.e., $w_{*}<-\frac{1}{3}$) are also
summarized in the same table. For convenience, we introduce the
parameter
\begin{equation}
\lambda_{s}=\{[\alpha
(1-c_{1}+c_{2})+(1+\alpha)(1-c_{1})A]^2+2(1-c_{1})(c_{1}+c_{2})[\alpha
(1-c_{1}+c_{2})+(1+\alpha)(1-c_{1})A]+[(1-c_{1})(c_{1}-c_{2})]^2\}^{\frac{1}{2}}.
\end{equation}
From Table I, we can clearly see that points $(a)$ and $(b)$ are not
stable if they exist, and point $(c)$ is stable under conditions
$c_{1}<1,c_{2}\geq 0$ and
$-\frac{\alpha}{1+\alpha}\frac{1-c_{1}+c_{2}}{1-c_{1}}\leq
A<\frac{1}{1+\alpha}(1-\alpha\frac{1-c_{1}+c_{2}}{1-c_{1}})$.
Furthermore, the stable attractor is a scaling solution since the
energy density of the MCG remains proportional to that of cold dark
matter, $\frac{\rho_{g*}}{\rho_{m*}}=\frac{1-c_{1}}{c_{2}}$, when
$c_{2}\neq0$. Thus, point $(c)$ is an accelerated scaling solution
that probably alleviates the coincidence problem. Additionally, for
the stable attractor point $(c)$, the location of the point depends
only on the coupling constant $c_{1}$ and $c_{2}$. The EoS of the
total cosmic fluid reads $w_{*}=-1$, which indicates that point
$(c)$ is a accelerated attractor ($w_{*}<-\frac{1}{3}$) and implies
that the universe will enter to a de sitter phase and accelerate
forever, and therefore, there is no singularity in the finite future
for any parameters. However, the EoS of the MCG is expressed by
$w_{g*}=-\frac{1-c_{1}+c_{2}}{1-c_{1}}=-1-\frac{c_{2}}{1-c_{1}}$,
that is to say, point $(c)$ has a phantom equation of state
($w_{g*}<-1$) for $c_{1}<1,c_{2}\geq 0$. According to Eq. (17), we
can obtain that $q_{*}=-1+\frac{3}{2}(1+y_{*})=-1$ for the stable
attractor point $(c)$, which also implies that the universe will
enter to a de sitter phase and accelerate forever.

\begin{figure}[tbp]
\includegraphics[width=0.3\textwidth]{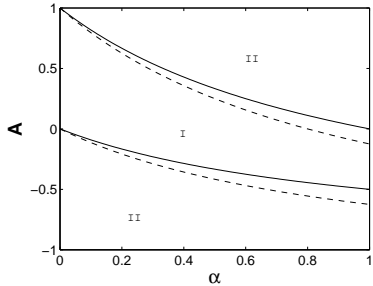}
\caption{The stable regions in the $(\alpha, A)$ parameter space.
The solid and dotted lines respectively denote the region for
$c_{1}=c_{2}=0$ (the modified Chaplygin gas model without
interaction) or $c_{1}\neq0,c_{2}=0$ (coupling between dark sectors
only proportional to the energy density of dark matter) and
$c_{1}=c_{2}=0.2$. In the region I (the region between the two same
lines), point $(c)$ is a stable accelerated attractor. The region II
represents the region of solution without physical meaning. }
\label{Fig.1}
\end{figure}

\begin{figure}[tbp]
\includegraphics[width=0.3\textwidth]{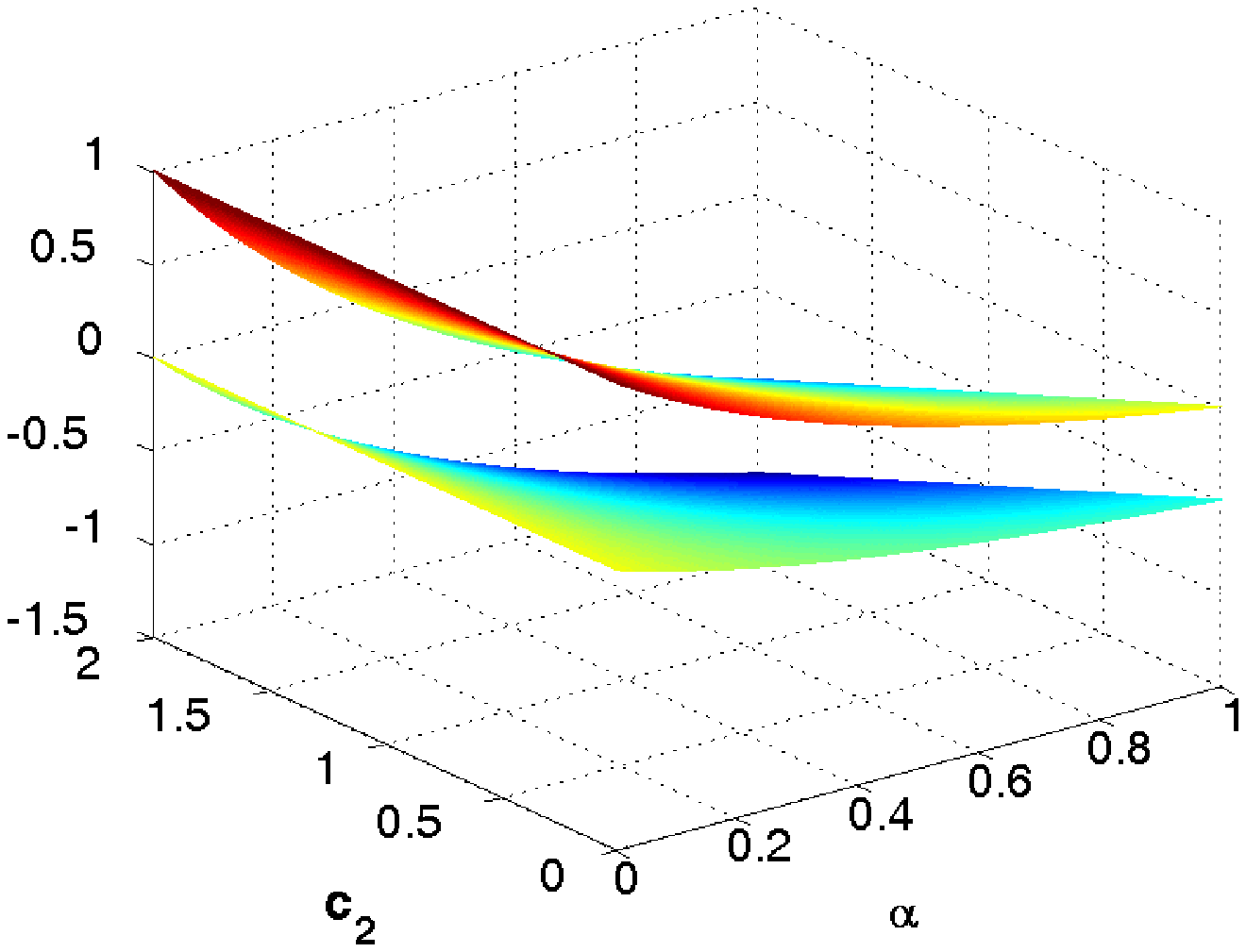}%
\includegraphics[width=0.3\textwidth]{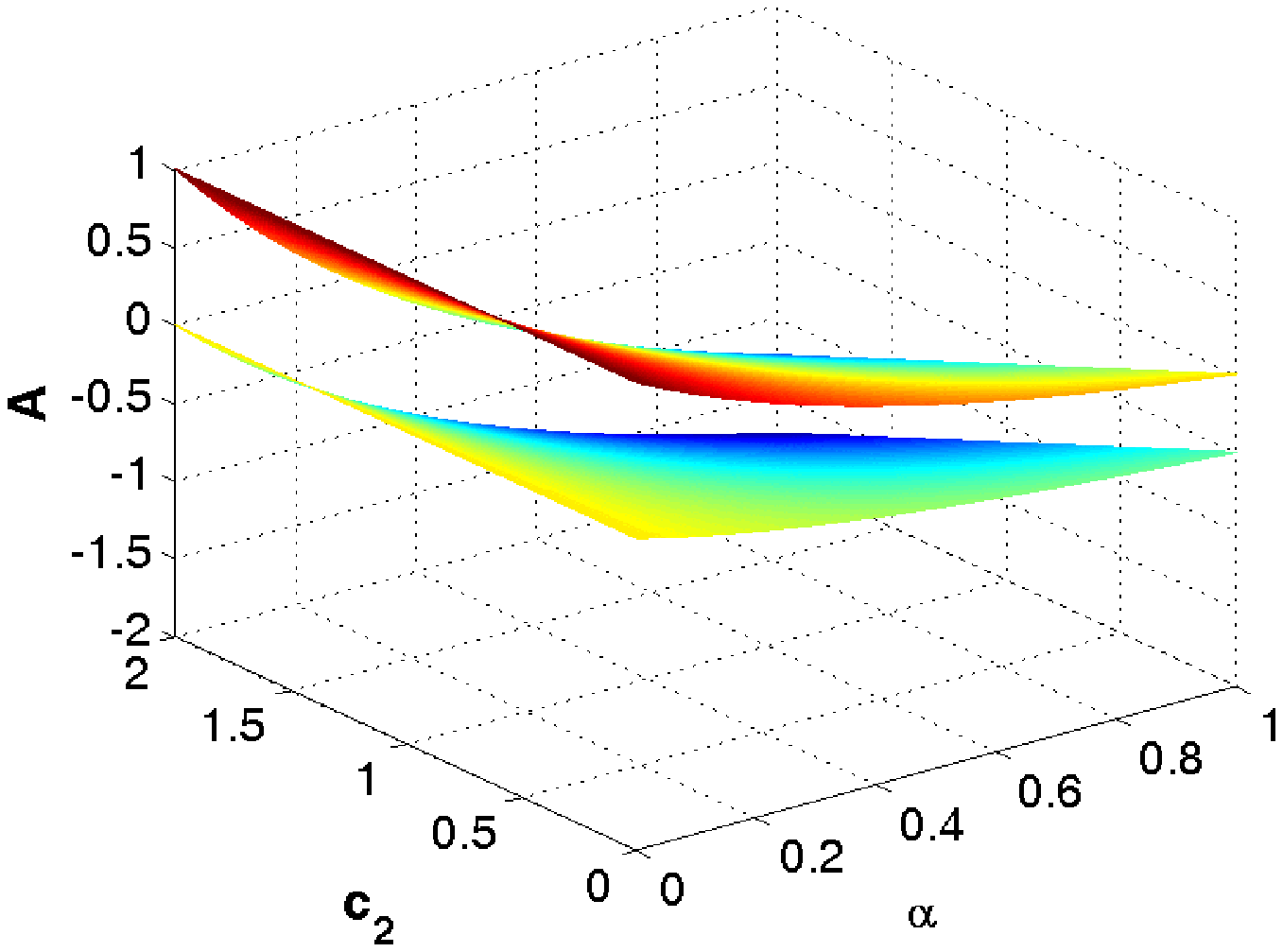}%
\includegraphics[width=0.3\textwidth]{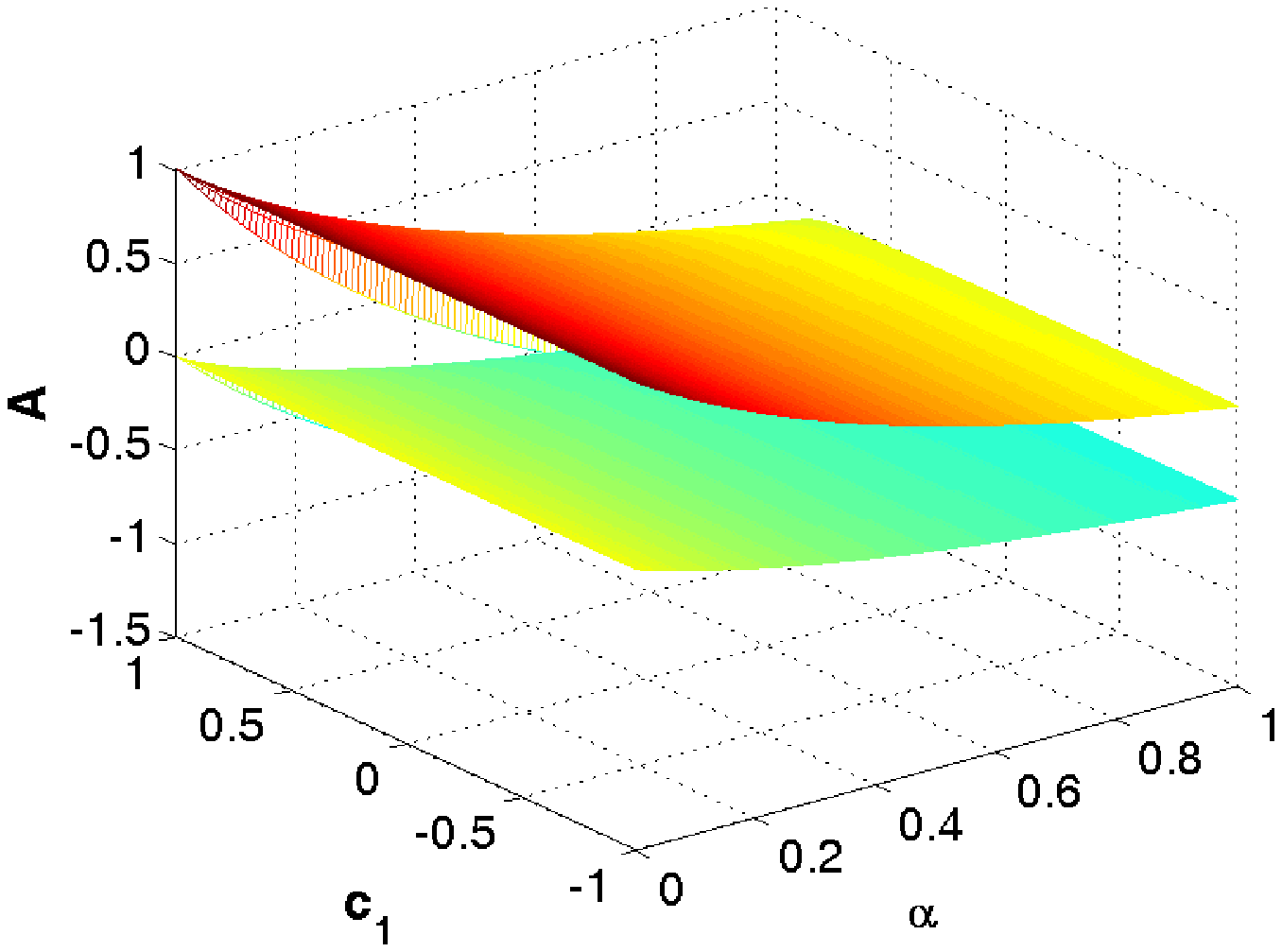}
\caption{The stable regions in the parameter space for the fixed
$c_{1}$ or $c_{2}$. The left two correspond to $c_{1}=0$ (coupling
between dark sectors only proportional to the energy density of dark
energy) and $c_{1}=0.4$, respectively. The right is for
$c_{2}=0.02$. The region between the two colorful planes represents
that point $(c)$ is a stable accelerated attractor. The other
regions correspond to the regions of solution without physical
meaning. } \label{Fig.2}
\end{figure}

\begin{figure}[tbp]
\includegraphics[width=0.3\textwidth]{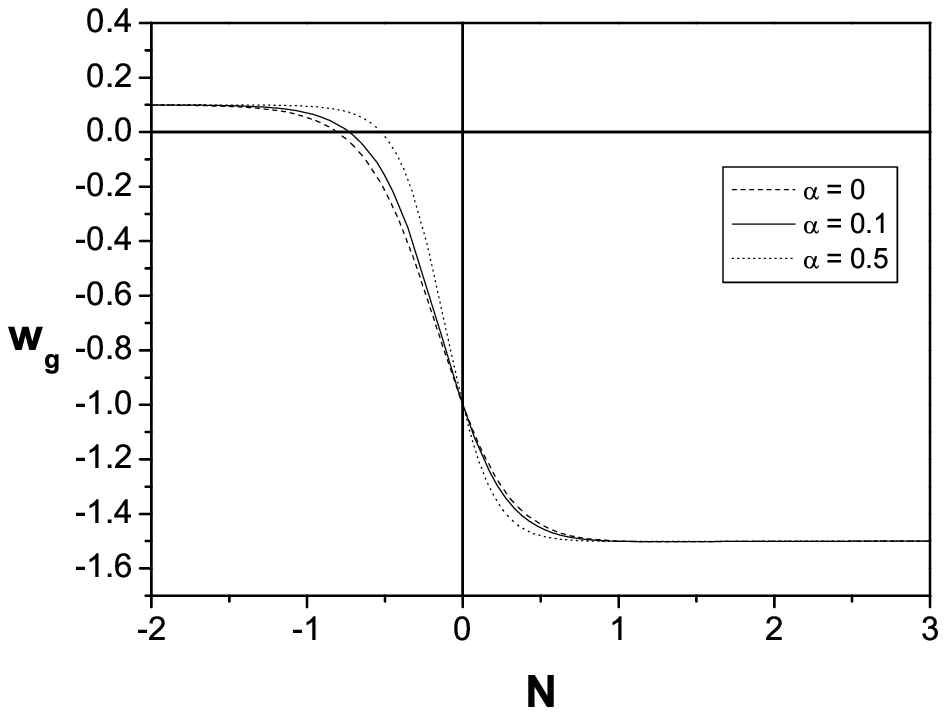}%
\includegraphics[width=0.3\textwidth]{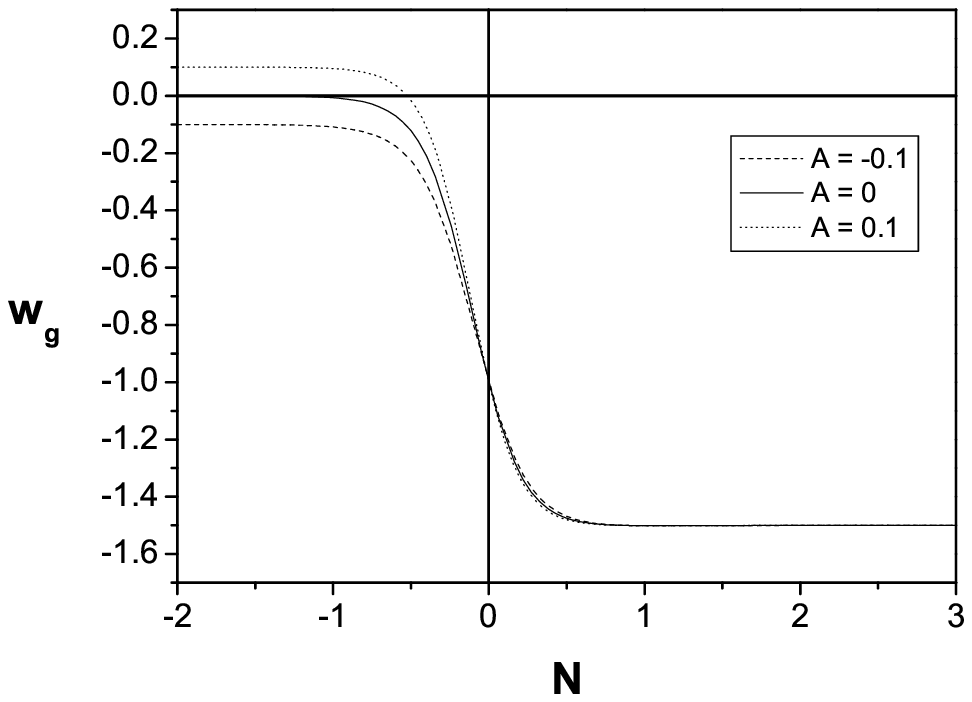}\\
\includegraphics[width=0.3\textwidth]{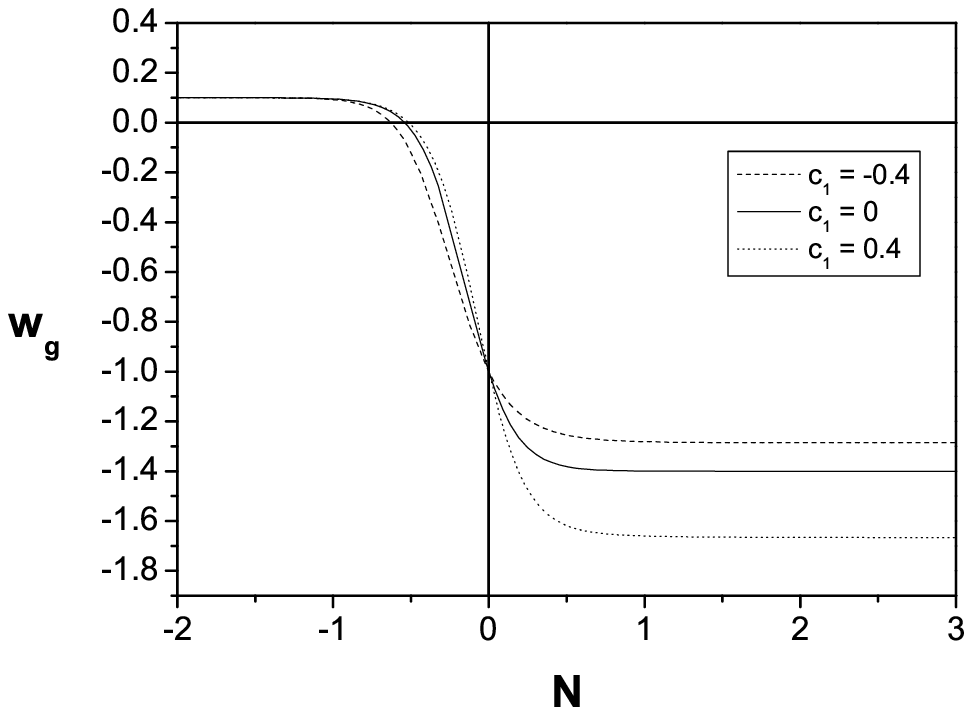}%
\includegraphics[width=0.3\textwidth]{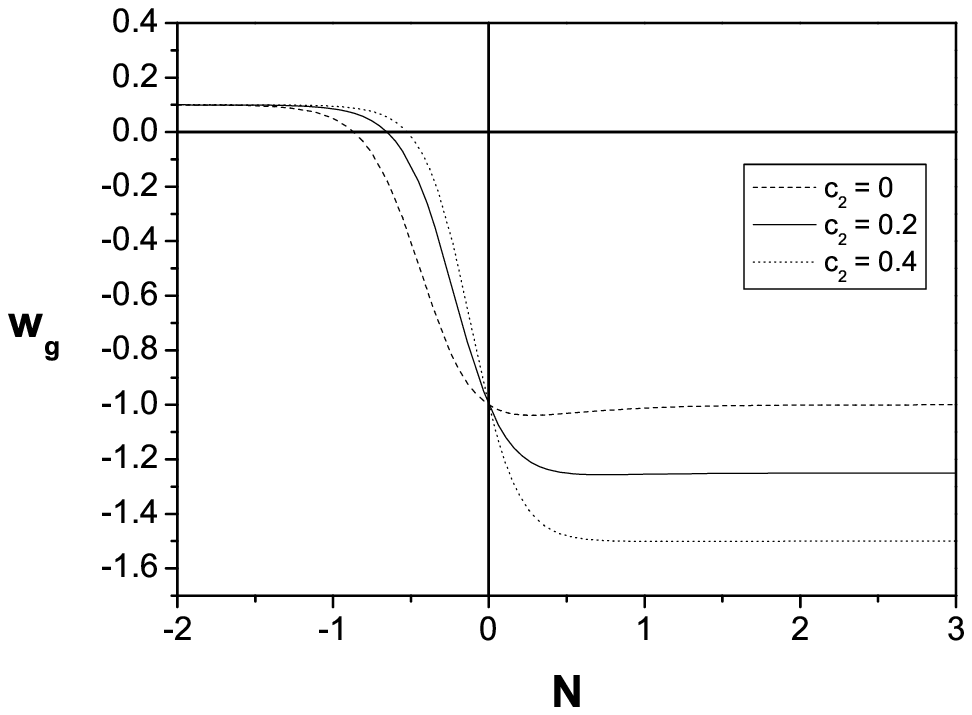}
\caption{The evolution of the EoSs of the modified Chaplygin gas
$w_{g}$ for different values of various parameters. The top two
plots correspond to the cases for different values of the modified
Chaplygin gas parameters $\alpha$ or $A$, when $c_{1}= 0.2$ and
$c_{2}=0.4$. In addition, we respectively fix the parameter $A=0.1$
and $\alpha=0.5$ in these two plots. The bottom two plots represent
the case for different values of the coupling constants $c_{1}$ or
$c_{2}$, when $\alpha=0.5$ and $A=0.1$. Moreover, in these two
plots, the parameter $c_{2}$ and $c_{1}$ are fixed to be $c_{2}=0.4$
and $c_{1}=0.2$ respectively.} \label{Fig.3}
\end{figure}

\subsection{B.~~~~Numerical Results}

In what follows, we numerically study the dynamical results from the
autonomous system (18)-(20) to clearly confirm the complicated
stability condition for the stable accelerated attractor point
$(c)$. In Figs.1 and 2, we depict the parameter space for point
$(c)$ to be stable. We suppose that $c_{1}=c_{2}=c$ or
$c_{1}\neq0,c_{2}=0$ in Fig.1. From the figure, we see that the
parameter space is independent of the coupling constant $c_{1}$ when
$c_{2}=0$, namely, the stable region of the modified Chaplygin gas
without interaction ($c_{1}=c_{2}=0$) is the same as that of the
interacting MCG, in which the interaction term is solely
proportional to the energy density of dark matter
($c_{1}\neq0,c_{2}=0$). Fig.2 shows that the stable regions in the
parameter space for the fixed $c_{1}$ or $c_{2}$. In the figure, the
left two plots is respectively for $c_{1}=0$ and $c_{1}=0.4$,
meanwhile, the range of $c_{2}$ is specially chosen as $0\leq
c_{2}<2$ to present the results more transparently though
$c_{2}\geq0$ is permitted. The right plot is shown by choosing
$c_{2}=0.02$, and also the range of $c_{1}$ is selected as $-1\leq
c_{1}<1$ for clarity although its allowed range is $c_{1}<1$.

\begin{figure}[tbp]
\includegraphics[width=0.3\textwidth]{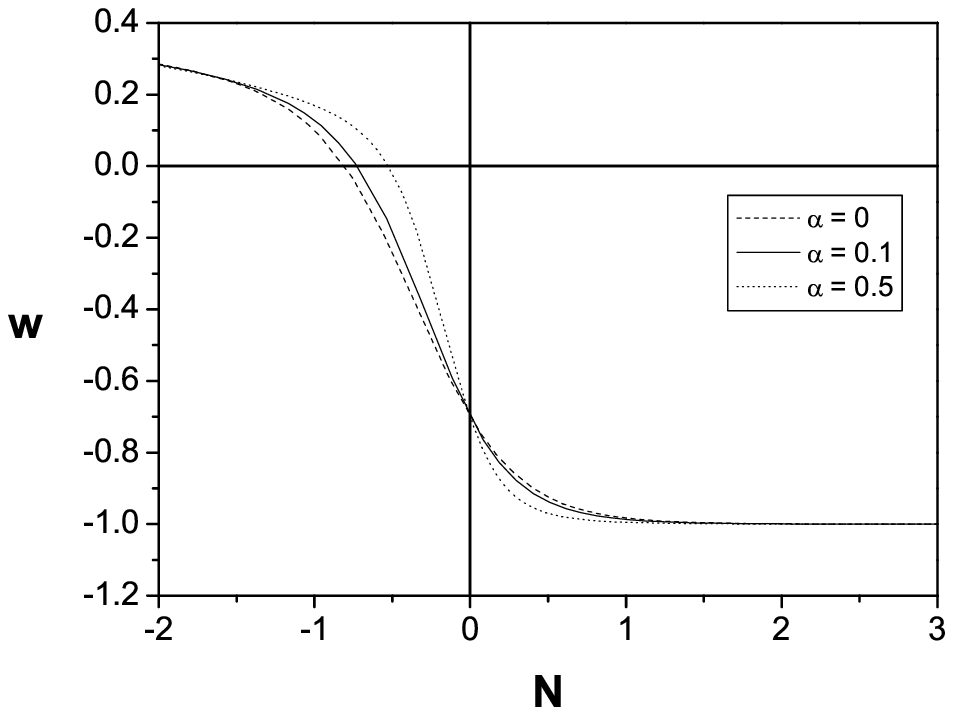}%
\includegraphics[width=0.3\textwidth]{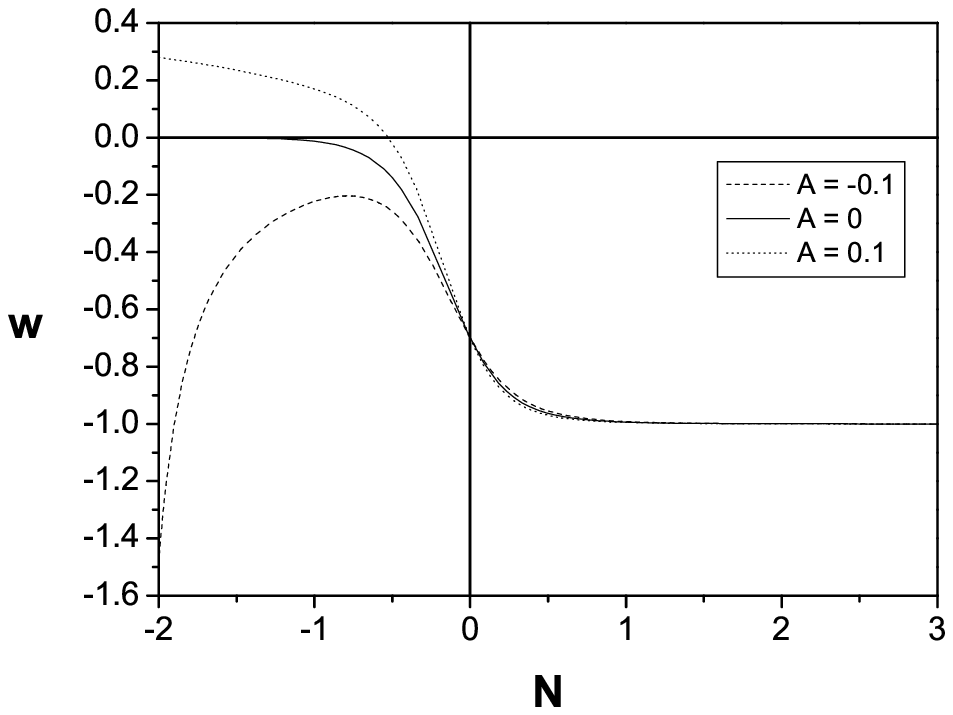}\\
\includegraphics[width=0.3\textwidth]{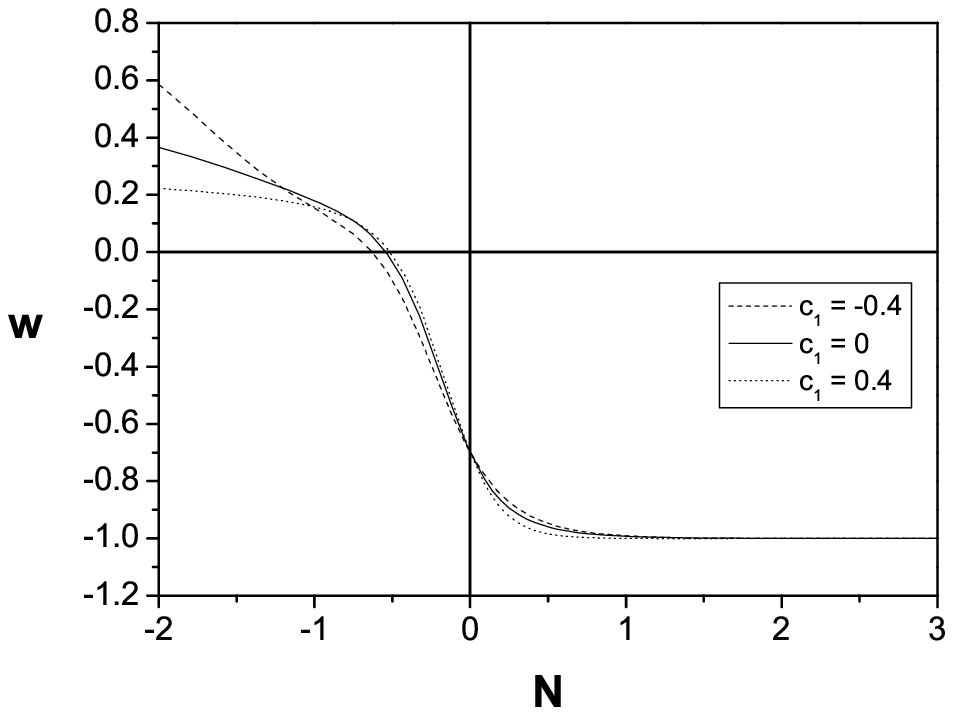}%
\includegraphics[width=0.3\textwidth]{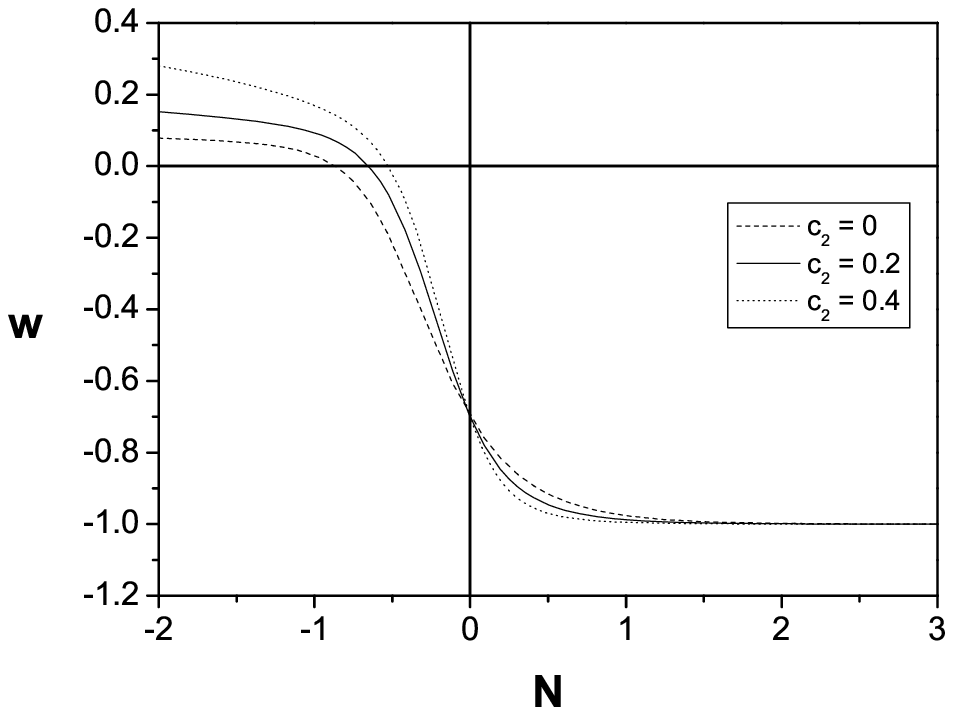}
\caption{The evolution of the EoSs of the total cosmic fluid $w$ for
different values of various parameters. In these four plots, we
respectively select the same parameters as those in Fig.3.}
\label{Fig.4}
\end{figure}

In Figs.3-4, we plot the evolution of the EoSs of the MCG $w_{g}$
and the total cosmic fluid $w$ for different parameters in the
stable region to illustrate the evolutional feature of our universe.
Meanwhile, the initial condition in these figures is taken to be
$x_{0}=\Omega_{g0}=0.7$, $y_{0}=-0.7$ (we assume $w_{g0}=-1$) and
$z_{0}=\Omega_{mo}=0.26$ to compare with the observation data. As
example, we choose several sets of values for the parameters
$\alpha$, $A$, $c_{1}$ and $c_{2}$ for clarity. From Fig.3, we see
that in the final state the EoS of the MCG, $w_{g}$, tends to a
constant, which is only determined by the coupling constants $c_{1}$
and $c_{2}$. Furthermore, the final EoS of the MCG, $w_{g*}$, is
always below $-1$ for different parameters in the stable region,
i.e., the MCG has a phantom equation of state in the final state. In
Fig.4, we find that the final EoS of the total cosmic fluid $w_{*}$
tends to $-1$, which is independent of any parameters in the stable
region. This indicates that the cosmic doomsday is avoided and the
universe accelerates forever. Note that the evolution of $w$ is
exotic when a set of parameters is chosen as $\alpha=0.5$, $A=-0.1$,
$c_{1}=0.2$ and $c_{2}=0.4$. In this case, the value of $w$ is
smaller than $-1$ at the high redshift, i.e., the interacting
modified Chaplygin gas model behave as phantom ($w<-1$) at the not
very early universe.

As the most significant parameter from the viewpoint of
observations, the deceleration parameter $q$ is also discussed. We
exhibit the evolution of the deceleration parameter $q$ for various
parameters in the stable region in Fig.5. The sets of values for all
parameters in this model are selected as those in Fig.3-4. From the
figure, we find that in the final state the deceleration parameter
$q$ tends to $-1$ for different values of various parameters. This
further indicates the universe enters to a de Sitter phase in the
final state. Moreover, varying any one of the MCG parameters
$\alpha$, $A$ and the coupling constants $c_{1}$, $c_{2}$, and
fixing the rest, we find that the larger the variational parameter
is, the smaller the transition redshift $z_{t}$ is, namely, the
later the transition from the deceleration phase to the acceleration
phase is. In addition, it is worthwhile to note that when we fix the
parameters $\alpha=0.5$, $c_{1}=0.2$ and $c_{2}=0.4$, and select
$A=-0.1$, there are two transition redshift: the first one denotes
the transition from the acceleration phase to the deceleration
phase; the second one represents the transition point from the
deceleration phase to the acceleration phase. This provides the
possibility that the evolution of our universe from acceleration to
deceleration and then from deceleration to acceleration.

\begin{figure}[tbp]
\includegraphics[width=0.3\textwidth]{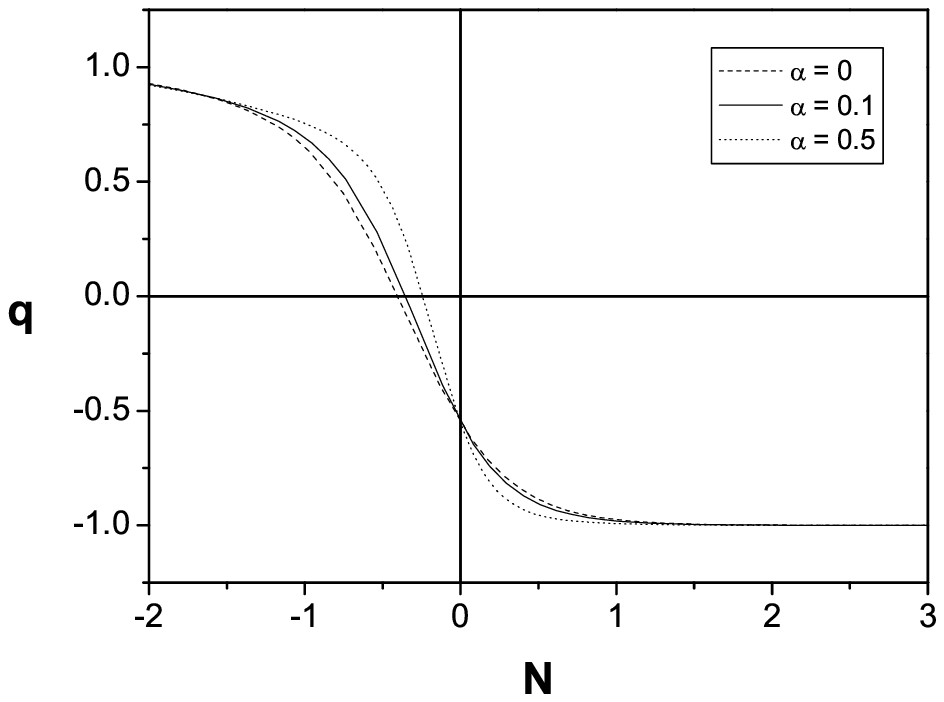}%
\includegraphics[width=0.3\textwidth]{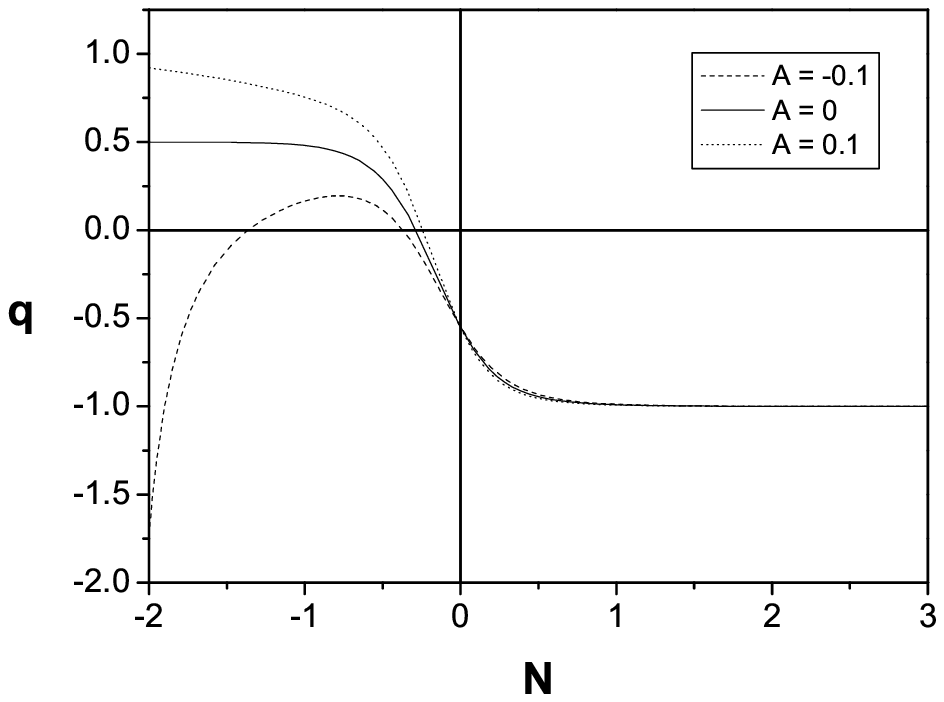}\\
\includegraphics[width=0.3\textwidth]{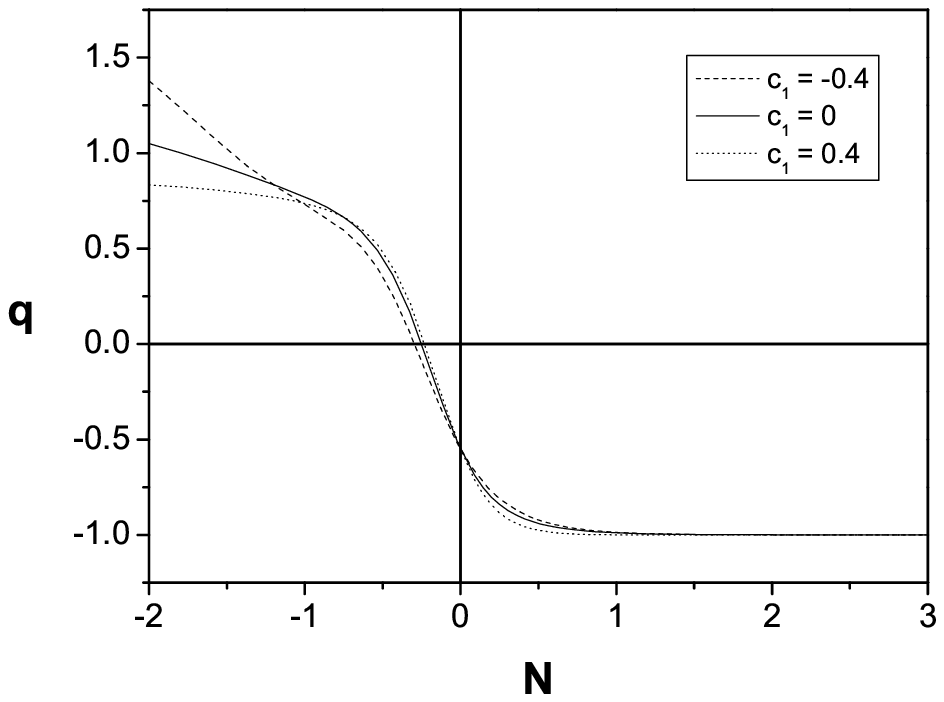}%
\includegraphics[width=0.3\textwidth]{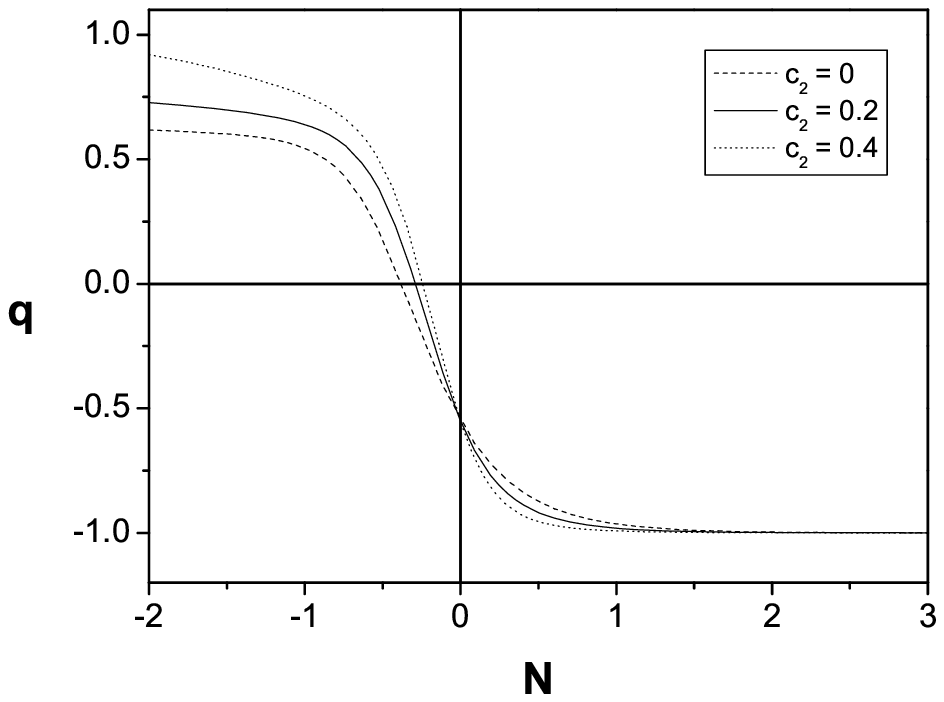}
\caption{The evolution of the deceleration parameter $q$ for
different values of various parameters. Also, the parameters in
these four plots are respectively chosen as those in Fig.3 for
clarity.} \label{Fig.5}
\end{figure}

\begin{figure}[tbp]
\includegraphics[width=0.3\textwidth]{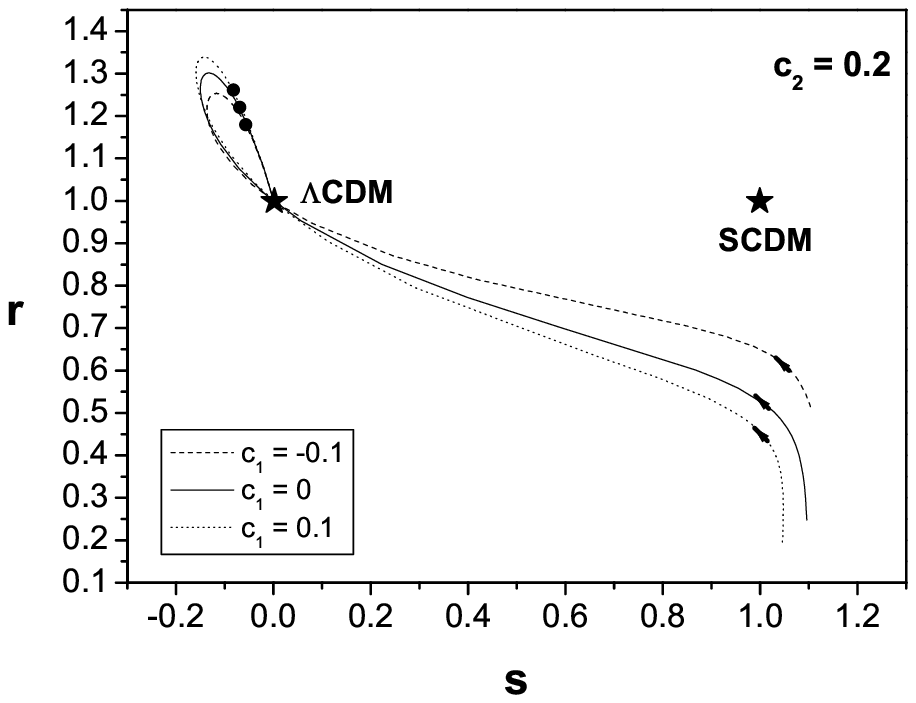}%
\includegraphics[width=0.3\textwidth]{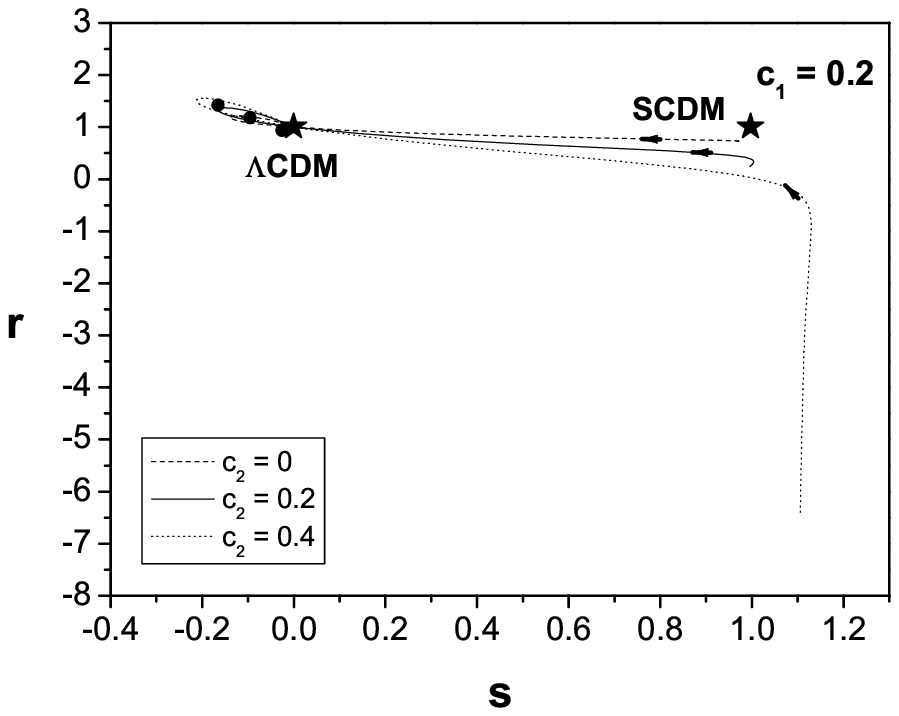}%
\includegraphics[width=0.3\textwidth]{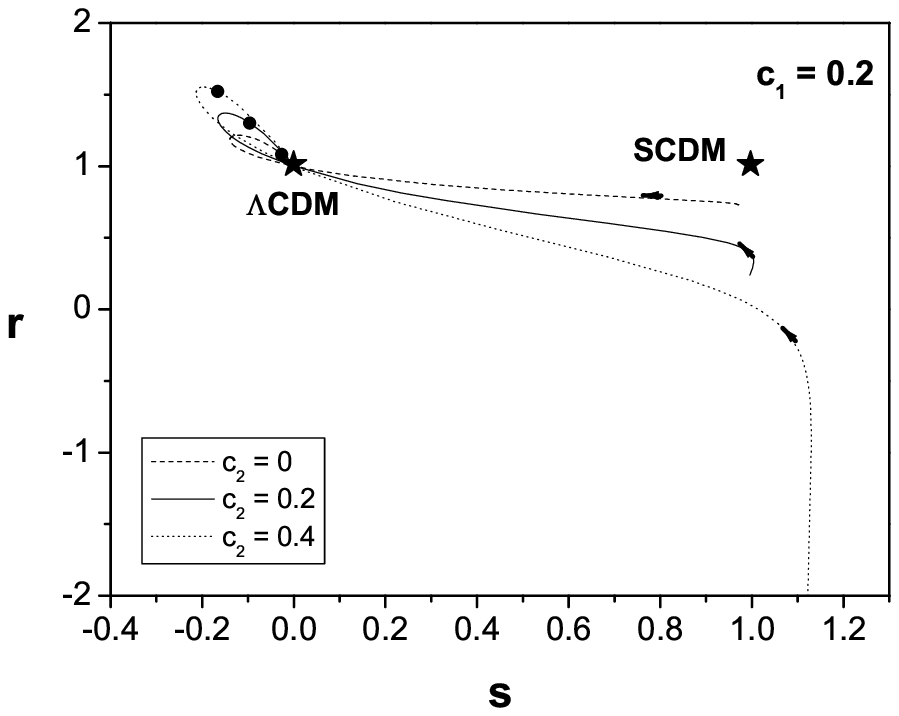}
\caption{The evolution of the statefinder for the interacting MCG
model in the $s-r$ plane with different coupling constants $c_{1}$
and $c_{2}$, which is illustrated in the left one and the right two
panels, respectively. To be clarity, we extract a part of the middle
panel to be the right one. In these three plots, we fix the
parameters $\alpha=0.5$ and $A=-0.1$. The solid points show the
current values of the statefinder parameters $(s_{0},r_{0})$. }
\label{Fig.6}
\end{figure}

\section{III.~~STATEFINDER DIAGNOSIS}

In this section, we study the dynamics of the interacting modified
Chaplygin gas model from the statefinder viewpoint. The so-called
statefinder parameter was first introduced by Sahni et
al.~\cite{phys44} in order to discriminate among more and more
cosmological models. It is constructed from the scalar $a$ and its
derivatives up to the third order. The statefinder pair $\{r,s\}$ is
defined as
\begin{equation}
r\equiv\frac{\dddot{a}}{aH^{3}},\mbox{}\hspace{15pt}s\equiv\frac{r-1}{3(q-1/2)}.
\end{equation}
Since different cosmological models exhibit qualitatively different
trajectories of evolution in the $s-r$ plane, the statefinder
parameter is a good tool to distinguish cosmological models. It has
a remarkable property for the basic spatially flat $\Lambda$CDM
model and the matter dominated universe SCDM, i.e., the statefinder
pair $\{r,s\}$ for $\Lambda$CDM model takes the constant value
$(0,1)$, while SCDM model corresponds to the fixed point $(1,1)$. We
can clearly identify the ``distance'' from a given cosmological
model to $\Lambda$CDM model in the $s-r$ plane, such as the
quintessence, the phantom, the Chaplygin gas, the holographic dark
energy models, the interacting dark energy models, and so forth,
which have been shown in the literatures~\cite{phys45}.

Generally, according to the reexpression of the deceleration
parameter $q$
\begin{equation}
q=-1-\frac{\dot{H}}{H},
\end{equation}
we can also rewrite the statefinder pair $\{r,s\}$ in terms of the
Hubble parameter $H$ and its first and second derivatives with
respect to the cosmic time $t$, $\dot{H}$ and $\ddot{H}$, as
\begin{eqnarray}
r&=&1+3\frac{\dot{H}}{H^2}+\frac{\ddot{H}}{H^3},\\
s&=-&\frac{3\frac{\dot{H}}{H^2}+\frac{\ddot{H}}{H^3}}{3(\frac{\dot{H}}{H^2}+\frac{3}{2})}.
\end{eqnarray}
Now we apply the statefinder parameter to the interacting MCG model.
By using Eqs. (6)-(11), the statefiner parameter can be concretely
expressed as
\begin{eqnarray}
r&=&1+\frac{9}{2}[(1+c_{2})x+y+c_{1}z][-\alpha\frac{y}{x}+(1+\alpha)A],\\
s&=&\frac{[(1+c_{2})x+y+c_{1}z][-\alpha\frac{y}{x}+(1+\alpha)A]}{y}.
\end{eqnarray}

The evolution of statefinder for the interacting MCG model in the
$s-r$ plane is plotted in Fig.6. The arrows in the figure denote the
evolution directions of the statefinder trajectories. In this
figure, we also use the initial condition: $x_{0}=\Omega_{g0}=0.7$,
$y_{0}=-0.7$ (we assume $w_{g0}=-1$) and $z_{0}=\Omega_{mo}=0.26$.
Since the evolution trajectory in the $s-r$ plane will be
interrupted when $y=0$ (i.e.,$w=0$), which is not our favorite case,
we carefully choose the parameter $A=-0.1$ based on the model
parameters selected before to avoid the interruption. This choice
can be seen from Fig.4, in which we can avoid the possibility of
$w=0$ during its evolution only if $A\geq0$ based on the model
parameters we selected. So in Fig.6, we take $A=-0.1$ and
$\alpha=0.5$ as example to investigate the effect of the coupling
constants $c_{1}$ and $c_{2}$, and furthermore, to discriminate
between the interacting MCG model and other dark energy models. From
Fig.6, we see that the evolution trajectories not only cross the
$\Lambda$CDM fixed point but also end at the fixed point for various
values of $c_{1}$ and $c_{2}$. However, the $r(s)$ curve could not
traverse the SCDM fixed point for any case. In addition, in the left
plot, the $r(s)$ curves are very different with various $c_{1}$ and
the fixed $c_{2}$, especially, the current values of $(s,r)$ are
different for various values of $c_{1}$. Also, we can clearly see
the tremendous difference between the current values $(s_{0},r_{0})$
and the $\Lambda$CDM fixed point, and the larger the value of
$c_{1}$ is, the greater the ``distance'' from the current value
$(s_{0},r_{0})$ to the fixed point is. On the other hand, fixing the
coupling constant $c_{1}$ and varying $c_{2}$ as illustrated in the
right two plots of Fig.6, we find that the $r(s)$ curves are very
different from each other with various $c_{2}$, and the current
values of $(s,r)$ are distinct for different values of $c_{2}$.
Also, we can clearly see the difference between the current values
$(s_{0},r_{0})$ and the $\Lambda$CDM fixed point, and the larger the
value of $c_{2}$ is, the greater the ``distance'' from
$(s_{0},r_{0})$ to the fixed point is. Therefore, we can conclude
that the statefinder diagnosis can not only discriminate the
interacting MCG model with different coupling constant but also
distinguish the interacting MCG model from other dark energy models.

\section{IV.~~Conclusions}

In previous sections, we have studied some physical properties of
the interacting MCG model. By considering an interaction term
between the MCG and cold dark matter, we study the dynamical
evolution of this model and pay our attention to the final state of
the universe. By our analysis, there exists a stable scaling
solution, which is characterized by a constant ratio of the energy
densities of the MCG and dark matter. This provides the possibility
to alleviate the coincidence problem. Furthermore, we see that the
final state is determined by the parameters of the MCG $\alpha$, $A$
and the coupling constants $c_{1}$ and $c_{2}$.

Interestingly, we find that the EoS of the MCG, $w_{g}$, tends to a
constant, which is only determined by the coupling constants $c_{1}$
and $c_{2}$. But both the EoS of the total cosmic fluid $w$ and the
deceleration parameter $q$ tend to $-1$, which are independent of
the choice of values for the parameters. This indicates that the
cosmic doomsday is avoided and the universe enters to a de Sitter
phase and thus accelerates forever. The transition from the
deceleration phase to the acceleration phase also depends on the MCG
parameters $\alpha$, $A$ and the coupling constants $c_{1}$,
$c_{2}$. Varying one of these four parameters and fixing the rest,
we find that the the larger the variational parameter is, the
smaller the transition redshift $z_{t}$ is, namely, the later the
transition from the deceleration phase to the acceleration phase is.
Further, we study the effect of the coupling constants $c_{1}$ and
$c_{2}$ on the dynamical evolution of this model from the
statefinder viewpoint. We clearly see that the coupling constants
$c_{1}$ and $c_{2}$ play a significant role during the dynamics of
the interacting MCG model. The evolution trajectories of the
statefinder for the interacting MCG model not only cross the
$\Lambda$CDM fixed point but also end at the fixed point for various
values of $c_{1}$ and $c_{2}$. However, the $r(s)$ curve could not
traverse the SCDM fixed point for any case. Moreover, fixing one of
the coupling constants $c_{1}$ and $c_{2}$, and varying the other
one, we find that the larger the variational parameter is, the
greater the ``distance'' from the current value $(s_{0},r_{0})$ to
the fixed point is. Thus, the statefinder parameter can discriminate
the interacting MCG model with different coupling constant. It is
also worthwhile to note that, we can distinguish this interacting
model from other dark energy models in the $s-r$ plane.

\section{Acknowledgements}

This work is a part of project 10675019 supported by NSFC.

\end{document}